\documentclass[fleqn]{mnras}
\usepackage{epsfig}
\usepackage{threeparttable}
\usepackage{multicol}
\usepackage{aas_macros,graphicx,times,multirow,amsmath,amssymb,color,longtable}
\usepackage[T1]{fontenc}
\usepackage{pdflscape}

\def\ltsima{$\; \buildrel < \over \sim \;$}
\def\lsim{\lower.5ex\hbox{\ltsima}}
\def\gtsima{$\; \buildrel > \over \sim \;$}
\def\gsim{\lower.5ex\hbox{\gtsima}}

\newcommand{\be}{\begin{equation}}
\newcommand{\en}{\end{equation}}

\def\nh{\hbox{$N_{\rm H}$}}
\def\flux {\mbox{erg cm$^{-2}$ s$^{-1}$}}
\def\lum {\mbox{erg s$^{-1}$}}

\def\aa {1E\,1547$-$5408}

\def\xte{XTE\,J1810$-$197}

\def\sgrd{SGR\,1627$-$41}

\def\lowba{SGR\,0418$+$5729}

\def\lowbb{Swift\,J1822.3$-$1606}
\def\galcen{SGR\,J1745$-$2900}
\def\sgras{Sgr\,A$^{\star}$}

\newcommand\cxo{{\em Chandra}}

\newcommand\xmm{{\em XMM--Newton}}
\newcommand{\nustar}{{\em NuSTAR}}

\newcommand{\swift}{{\em Swift}}
\newcommand{\fermi}{{\em Fermi}}

\begin{document}

\title[Chandra campaign of SGR\,J1745$-$2900 outburst decay]{\cxo\ monitoring of the Galactic Centre magnetar SGR\,J1745$-$2900 during the initial 3.5 years of outburst decay}
\author[F. Coti Zelati et al.]
{F. Coti Zelati,$^{1,2,3,4}$\thanks{E-mail: cotizelati@ice.csic.es} N. Rea,$^{2,4}$ R. Turolla,$^{5,6}$ J. A. Pons,$^{7}$ A. Papitto,$^{8}$ P. Esposito,$^{2}$ \newauthor G.~L. Israel,$^{8}$ S. Campana,$^{3}$ S. Zane,$^{6}$
A. Tiengo,$^{9,10,11}$ R.~P. Mignani,$^{9,12}$ S. Mereghetti,$^{9}$ \newauthor F.~K. Baganoff,$^{13}$ D. Haggard,$^{14,15}$  G. Ponti,$^{16}$  D.~F. Torres,$^{4,17}$  A. Borghese$^{2}$ and J. Elfritz$^{2}$\\
$^{1}$ Universit\`a dell'Insubria, via Valleggio 11, I-22100 Como, Italy\\
$^{2}$ Anton Pannekoek Institute for Astronomy, University of Amsterdam, Postbus 94249,  NL-1090-GE Amsterdam, The Netherlands\\
$^{3}$ INAF -- Osservatorio Astronomico di Brera, via Bianchi 46, I-23807 Merate (LC), Italy\\
$^{4}$ Institute of Space Sciences (ICE, CSIC--IEEC), Carrer de Can Magrans, S/N, 08193, Barcelona, Spain\\
$^{5}$ Dipartimento di Fisica e Astronomia, Universit\`a di Padova, via F. Marzolo 8, I-35131 Padova, Italy\\
$^{6}$ Mullard Space Science Laboratory, University College London, Holmbury St. Mary, Dorking, Surrey RH5 6NT, UK\\
$^{7}$ Departament de Fisica Aplicada, Universitat d'Alacant, Ap. Correus 99, E-03080 Alacant, Spain\\
$^{8}$ INAF -- Osservatorio Astronomico di Roma, via Frascati 33, I-00040 Monteporzio Catone, Roma, Italy\\
$^{9}$ INAF -- Istituto di Astrofisica Spaziale e Fisica Cosmica, via E. Bassini 15, I-20133 Milano, Italy\\
$^{10}$ Istituto Universitario di Studi Superiori, piazza della Vittoria 15, I-27100 Pavia, Italy\\
$^{11}$ Istituto Nazionale di Fisica Nucleare, Sezione di Pavia, via A. Bassi 6, I-27100 Pavia, Italy\\
$^{12}$ Kepler Institute of Astronomy, University of Zielona G\'{o}ra, Lubuska 2, 65-265, Zielona G\'ora, Poland\\
$^{13}$ Kavli Institute for Astrophysics and Space Research, Massachusetts Institute of Technology, Cambridge, MA 02139, USA\\
$^{14}$ Department of Physics, McGill University, 3600 University St., Montreal, QC H3A 2T8, Canada\\
$^{15}$ McGill Space Institute, McGill University, Montreal, QC H3A 2A7, Canada\\
$^{16}$ Max Planck Institut f\"ur Extraterrestriche Physik, Giessenbachstrasse, D-85748 Garching, Germany\\
$^{17}$ Instituci\'o Catalana de Recerca i Estudis Avan\c{c}ats (ICREA) Barcelona, Spain\\
}

\date{}
\maketitle

\begin{abstract}
We report on 3.5 years of \cxo\, monitoring of the Galactic Centre magnetar \galcen\ since its outburst onset in April 2013. The magnetar spin-down has shown at least two episodes of period derivative increases so far, and it has slowed down regularly in the past year or so. We observed a slightly increasing trend in the time evolution of the pulsed fraction, up to $\sim 55$ per cent in the most recent observations. \galcen\ has not reached the quiescent level yet, and so far the overall outburst evolution can be interpreted in terms of a cooling hot region on the star surface. We discuss possible scenarios, showing in particular how the presence of a shrinking hot spot in this source is hardly reconcilable with internal crustal cooling and favours the untwisting bundle model for this outburst. Moreover, we also show how the emission from a single uniform hot spot is incompatible with the observed pulsed fraction evolution for any pair of viewing angles, suggesting an anisotropic emission pattern.
\end{abstract}

\begin{keywords}
methods: data analysis, observational -- Galaxy: centre -- stars: magnetars -- X-rays: individual: SGR\,J1745-2900
\end{keywords}

\section{Introduction}

One of the most enigmatic classes of the neutron star population comprises 23 objects whose observational manifestations can be explained only by exceptionally large magnetic fields, up to $\sim10^{14}-10^{15}$~G at the surface (see Olausen \& Kaspi 2014 and the McGill magnetar Catalog\footnote{\url{http://www.physics.mcgill.ca/~pulsar/magnetar/main.html}.}). These neutron stars are commonly referred to as magnetars (see e.g. Thompson \& Duncan 1993, 1995). In the past few years, new breakthroughs have clearly demonstrated how these strong fields might be hidden close or inside the neutron star crust (Rea et al. 2010; Turolla et al. 2011; Tiengo et al. 2013).

Magnetars might release their large magnetic energy via steady X-ray emission (L$_X\sim10^{32}-10^{36}$~\lum), or through bright flares and outbursts occurring at unpredictable times with significantly different timescales (milliseconds to years) and energetics ($E\sim10^{38}-10^{45}$~erg; see Turolla, Zane \& Watts 2015 and Kaspi \& Beloborodov 2017 for recent reviews). The continuous surveys of the hard X-/gamma-ray sky by the Burst Alert Telescope (Barthelmy et al. 2005) aboard \swift\ and the Gamma-ray Burst Monitor (Meegan et al. 2009) onboard \fermi\ represent nowadays one of the most fruitful channels to spot new members of the class, yielding a discovery rate of almost one candidate per year through the detection of their flaring activity.
 
\galcen\ is located at an angular distance of only 2.4 arcsec from the 4-million-solar mass black hole at the centre of the Milky Way, Sagittarius~A* (\sgras\ hereafter). Its existence was heralded on 2013 April 25, following the detection of a $\sim$30-ms burst of soft gamma-rays (Kennea et al. 2013) and the discovery of a bright X-ray counterpart ($L_X \sim 5 \times 10^{35}$ erg~s$^{-1}$ for an assumed distance of 8.3 kpc; Genzel et al. 2010). Coherent pulsations at a period of 3.76 s were detected both in the X-ray and radio bands (Kennea et al. 2013; Mori et al. 2013; Rea et al. 2013a; Shannon \& Johnston 2013; Kaspi et al. 2014; Lynch et al. 2015; Pennucci et al. 2015), making \galcen\ the fourth confirmed radio-loud magnetar known to date after \xte, \aa\ and PSR\,1622$-$4950 (Camilo et al. 2006, 2007; Levin et al. 2010).

\galcen\ has been regularly observed by \cxo\ and \xmm\ since its discovery, as part of a monitoring program of the Galactic Centre primarily devoted to the investigation of the flaring activity of \sgras\ (e.g., Ponti et al. 2015 and references therein) and the effects of its interaction with the cold, dusty object G2. Analysis of the magnetar X-ray properties up to 2014 September unveiled an extremely slow decay of the X-ray flux (Coti Zelati et al. 2015), challenging the neutron star crustal cooling models successfully applied to other magnetar outburst decays. Although several observations have been performed with the European Photon Imaging Cameras (EPIC) on board \xmm\ over the last 3 years, \galcen\ was above the background level only during the first $\sim1.5$~yr of the outburst. This, and the presence of the nearby bright X-ray transient Swift\,J174540.7$-$290015 (discovered in 2016 February at only 16 arcsec from the magnetar and surrounded by a dust scattering halo; see Ponti et al. 2016), complicate detailed \xmm\ studies of the outburst decay of \galcen\ at late times\footnote{The full-width at half-maximum of the EPIC point spread function is about 6 arcsec.}. On the other hand, \cxo\ provides a prime opportunity to accurately investigate the evolution of the magnetar soft X-ray emission in the very long term, owing to the sub-arcsecond angular resolution of its X-ray instruments.

This paper presents 11 new \cxo\ observations of \galcen\ which, complemented with those we already described in our previous study (Coti Zelati et al. 2015), represent an unprecedented data set covering a time span of 3.5 years since the outburst onset. We include the \xmm\ data of the early phases of the outburst only to refine our timing solution, and refer to Coti Zelati et al. (2015) for details on the results of the spectral analysis. The paper is structured as follows: we describe the observations and the data reduction procedures in Section~\ref{obs}. We report on the timing and spectral analysis in Section~\ref{analysis}. Discussion of our results follows in Section~\ref{discussion}.

\begin{table*}
\caption{Journal of \cxo\ and \xmm/EPIC observations of \galcen. The 5-digit sequences refer to \cxo\ observations, the 10-digit sequences denote the \xmm\ observations.}
\begin{threeparttable}
\begin{tabular}{lclrcc}
\hline
Obs. ID     					& Mid point of observation		& Start time (TT)			& End time (TT)	& Exposure time	& 0.3--10 keV source net count rate\\
							& (MJD)					& \multicolumn{2}{c}{(yyyy/mm/dd hh:mm:ss)} 		& (ks)			& (counts s$^{-1}$)\\
\hline
14701\tnote{*}					& 56\,411.70	& 2013/04/29 15:14:12		& 2013/04/29 18:14:50			& 9.7				& $0.081 \pm 0.003$\\
14702						& 56\,424.55	& 2013/05/12 10:38:50		& 2013/05/12 15:35:56			& 13.7 			& $0.545 \pm 0.006$\\
15040\tnote{**}					& 56\,437.63	& 2013/05/25 11:38:37 		& 2013/05/25 18:50:50			& 23.8 			& $0.150 \pm 0.003$\\
14703						& 56\,447.48	& 2013/06/04 08:45:16		& 2013/06/04 14:29:15			& 16.8			& $0.455 \pm 0.005$\\
15651\tnote{**} 					& 56\,448.99	& 2013/06/05 21:32:38 		& 2013/06/06 01:50:11			& 13.8 			& $0.141 \pm 0.003$\\
15654\tnote{**}					& 56\,452.25	& 2013/06/09 04:26:16 		& 2013/06/09 07:38:28			& 9.0 			& $0.128 \pm 0.004$\\
14946						& 56\,475.41	& 2013/07/02 06:57:56		& 2013/07/02 12:46:18			& 18.2			& $0.392 \pm 0.005$\\
15041						& 56\,500.36	& 2013/07/27 01:27:17		& 2013/07/27 15:53:25			& 45.4			& $0.346 \pm 0.003$\\
15042						& 56\,516.25	& 2013/08/11 22:57:58		& 2013/08/12 13:07:47			& 45.7			& $0.317 \pm 0.003$\\
0724210201\tnote{***}			& 56\,535.19	& 2013/08/30 20:30:39		& 2013/08/31 12:28:26			& 55.6/57.2/57.2	& $0.657\pm0.004$\\
14945						& 56\,535.55	& 2013/08/31 10:12:46		& 2013/08/31 16:28:32			& 18.2			& $0.290 \pm 0.004$\\
0700980101\tnote{***}			& 56\,545.37	& 2013/09/10 03:18:13 		& 2013/09/10 14:15:07			& 35.7/37.3/37.3	& $0.636\pm0.005$\\
15043						& 56\,549.30	& 2013/09/14 00:04:52		& 2013/09/14 14:19:20			& 45.4			& $0.275 \pm 0.002$\\
14944						& 56\,555.42	& 2013/09/20 07:02:56		& 2013/09/20 13:18:10			& 18.2			& $0.273 \pm 0.004$\\
0724210501\tnote{***}			& 56\,558.15	& 2013/09/22 21:33:13		& 2013/09/23 09:26:52			& 41.0/42.6/42.5	& $0.607\pm0.005$\\
15044						& 56\,570.01	& 2013/10/04 17:24:48		& 2013/10/05 07:01:03			& 42.7			& $0.255 \pm 0.002$\\
14943						& 56\,582.78	& 2013/10/17 15:41:05		& 2013/10/17 21:43:58			& 18.2			& $0.246 \pm 0.004$\\
14704						& 56\,588.62	& 2013/10/23 08:54:30		& 2013/10/23 20:43:44			& 36.3			& $0.240 \pm 0.003$\\
15045						& 56\,593.91	& 2013/10/28 14:31:14		& 2013/10/29 05:01:24			& 45.4			& $0.234 \pm 0.002$\\
16508   						& 56\,709.77	& 2014/02/21 11:37:48  		& 2014/02/22 01:25:55			& 43.4			& $0.156 \pm 0.002$\\
16211 						& 56\,730.71	& 2014/03/14 10:18:27		& 2014/03/14 23:45:34			& 41.8			& $0.149 \pm 0.002$\\
0690441801\tnote{***}			& 56\,750.72	& 2014/04/03 05:23:24		& 2014/04/04 05:07:01			& 83.5/85.2/85.1	& $0.325\pm0.003$\\
16212						& 56\,751.40	& 2014/04/04 02:26:27		& 2014/04/04 16:49:26			& 45.4			& $0.135 \pm 0.002$\\
16213						& 56\,775.41	& 2014/04/28 02:45:05		& 2014/04/28 17:13:57			& 45.0			& $0.128 \pm 0.002$\\
16214						& 56\,797.31	& 2014/05/20 00:19:11		& 2014/05/20 14:49:18			& 45.4			& $0.118 \pm 0.002$\\
16210						& 56\,811.24	& 2014/06/03 02:59:23		& 2014/06/03 08:40:34			& 17.0			& $0.110 \pm 0.003$\\
16597						& 56\,842.98	& 2014/07/04 20:48:12		& 2014/07/05 02:21:32			& 16.5			& $0.097 \pm 0.002$\\
16215						& 56\,855.22	& 2014/07/16 22:43:52 		& 2014/07/17 11:49:38 			& 41.5			& $0.090 \pm 0.001$\\
16216						& 56\,871.43	& 2014/08/02 03:31:41 		& 2014/08/02 17:09:53			& 42.7 			& $0.085 \pm 0.001$\\
16217						& 56\,899.43	& 2014/08/30 04:50:12 		& 2014/08/30 15:45:44			& 34.5 			& $0.079 \pm 0.002$\\
0743630201\tnote{***}			& 56\,900.02 	& 2014/08/30 19:37:28 		& 2014/08/31 05:02:43			& 32.0/33.6/33.6	& $0.221\pm0.004$\\
0743630301\tnote{***}			& 56\,901.02	& 2014/08/31 20:40:57		& 2014/09/01 04:09:34			& 25.0/26.6/26.6	& $0.219\pm0.004$\\
0743630401\tnote{***}			& 56\,927.94	& 2014/09/27 17:47:50		& 2014/09/28 03:05:37			& 25.7/32.8/32.8	& $0.194\pm0.004$\\
0743630501\tnote{***}			& 56\,929.12	& 2014/09/28 21:19:11		& 2014/09/29 08:21:11			& 37.8/39.4/39.4	& $0.208\pm0.004$\\
\hline
\hline
16218\tnote{$\dagger$}			& 56\,950.59	& 2014/10/20 08:22:28		& 2014/10/20 19:59:16			& 36.3			& $0.071 \pm 0.001$\\	
16963\tnote{$\dagger$}			& 57\,066.18	& 2015/02/13 00:42:04		& 2015/02/13 08:09:46			& 22.7			& $0.056 \pm 0.002$\\
16966\tnote{$\dagger$}			& 57\,156.53	& 2015/05/14 08:46:51		& 2015/05/14 16:26:52			& 22.7			& $0.045  \pm 0.001$ \\
16965\tnote{$\dagger$}			& 57\,251.60	& 2015/08/17 10:35:47		& 2015/08/17 18:13:11			& 22.7 			& $0.035  \pm 0.001$\\
16964\tnote{$\dagger$}			& 57\,316.41	& 2015/10/21 06:04:57		& 2015/10/21 13:23:22			& 22.6			& $0.026  \pm 0.001$\\
18055\tnote{$\dagger$} 			& 57\,431.53	& 2016/02/13 08:59:23		& 2016/02/13 16:26:00			& 22.7			& $0.0133  \pm 0.0008$ \\
18056\tnote{$\dagger$}  			& 57\,432.76	& 2016/02/14 14:46:01		& 2016/02/14 21:44:19			& 21.8			& $0.0146  \pm 0.0009$\\
18731\tnote{$\dagger$} 			& 57\,582.27	& 2016/07/12 18:23:59		& 2016/07/13 18:42:51			& 78.4			& $0.0112  \pm 0.0004$\\
18732\tnote{$\dagger$}			& 57\,588.00	& 2016/07/18 12:01:38		& 2016/07/19 12:09:00			& 76.6			& $0.0118  \pm 0.0004$\\
18057\tnote{$\dagger$}			& 57\,669.95	& 2016/10/08 19:07:12		& 2016/10/09 02:38:59			& 22.7			& $0.0123 \pm 0.0008$ \\
18058\tnote{$\dagger$}			& 57\,675.61	& 2016/10/14 10:47:43		& 2016/10/14 18:16:44			& 22.7  			& $0.0122 \pm 0.0007$ \\		
\hline
\end{tabular}
\begin{tablenotes}
\item[*] \cxo/HRC observation.
\item[**] \cxo/ACIS-S grating observations.
\item[***] \xmm\ observations. Exposure times are reported for the pn, MOS1 and MOS2 cameras. Source net count rates refer to the pn detector.
\item[$\dagger$] New unpublished \cxo\ observations.
\end{tablenotes}
\end{threeparttable}
\label{tab:log}
\end{table*}

\section{Observations and data extraction}
\label{obs}

The {\em Chandra X-ray Observatory} observed \galcen\ 37 times between 2013 April 29 and 2016 October 14, for a total dead-time corrected on-source exposure time of about 1.17~Ms (see Table~\ref{tab:log} for an updated journal of the observations). Except for the first pointing (ID 14701), which was carried out with the spectroscopic detector of the High Resolution Camera (HRC-S; Zombeck et al. 1995), all observations were performed with the Advanced CCD Imaging Spectrometer spectroscopic array (ACIS-S; Garmire et al. 2003) operated in timed-exposure imaging mode and with faint telemetry format. A 1/8 sub-array was adopted to achieve a time resolution of 0.44104~s and allow a proper characterization of the pulsations at the magnetar spin period, $\sim 3.76$~ s. The source was always positioned on the back-illuminated S3 chip.  

The new data were processed and analysed using the \cxo\ Interactive Analysis of Observations software (\textsc{ciao}, v. 4.8; Fruscione et al. 2006) and the most recent version of the calibration files (\textsc{caldb}, v. 4.7.2). As it was done in our previous study (Coti Zelati et al. 2015), source photons were collected within a 1.5-arcsec circle centered on the source position. For each observation the background was estimated against many regions significantly differing in shape, size and proximity to the source. A 1.5-arcsec circle at the target position in archival (i.e., pre-outburst) ACIS-S observations of the field was also adopted to gauge the background level. All light curves were visually inspected and filtered for flares from particle-induced background (e.g., Markevitch et al. 2003). The source net count rate further decreased in the last 11 observations between 2014 October 20 and 2016 October 14 (see Table \ref{tab:log}), and pile-up did not affect any of these data sets, as also confirmed by applying the \textsc{pileup$_{-}$map} tool on the event files. All analyses were restricted to photons having energies between 0.3 and 8 keV. Photon arrival times were referred to the Solar system barycentre reference frame using the task \textsc{axbary}. Source and background spectra, redistribution matrices and ancillary response files were generated via \textsc{specextract}. Background-subtracted spectra were then grouped to have at least 50 counts in each energy channel, to enable the use of the $\chi^2$ statistics to assess the goodness-of-fit. All uncertainties on the parameters are quoted at the 1-$\sigma$ confidence level for a single parameter of interest, unless otherwise specified.

\galcen\ was observed 6 more times between 2015 April 25 and 2016 July 24 with the ACIS imaging array (ACIS-I) aboard \cxo. In all these cases \galcen\ lied at an off-axis angle of about 9.8 arcmin from the aim point (the \sgras\ complex). The point spread function of the ACIS detectors is known to exhibit significant variations in size and shape across the focal plane, and at such off-axis angles the \textsc{ciao} tool \textsc{psfsize$_-$srcs} yields indeed an estimate of about 6 and 10 arcsec for the radii enclosing 50 and 90 per cent of the source photons, respectively (at an energy of 3~keV). Such large extraction regions encompass the X-ray counterpart of \sgras, and would certainly result in unreliable estimates of the magnetar flux. Moreover, the coarse time resolution achieved with the timed-exposure configuration of ACIS-I (Nyquist limit of about 6.48~s) precludes the detection of the pulsations. We hence decided not to include these observations in our analysis.

\begin{table*}
\caption{Timing solutions. The first solution (MJD 56\,411.6 -- 56\,475.5) is taken from Rea et al. (2013a), the second (MJD 56\,500.1 -- 56\,594.2) corresponds to Solution A by Coti Zelati et al. (2015), and the third is reported in Section~\ref{timing} of this work and represents an updated solution of Solution B by Coti Zelati et al. (2015) over a longer temporal baseline. Uncertainties were evaluated at the 1$\sigma$ confidence level, scaling the uncertainties by the value of the rms ($\sqrt{\chi_\nu^2}$) of the respective fit to account for the presence of unfitted residuals.}
\begin{tabular}{lccc}
\hline \vspace{0.1cm}	
Validity range [ MJD] 	                 	&  56\,411.6 -- 56\,475.5		& 56\,500.1 -- 56\,594.2           		& 56\,709.5 -- 57\,588.5			\\ \vspace{0.1cm}
Epoch $T_0$ [ MJD]    	        			& 56\,424.5509871 			& 56\,513.0                          		& 56\,710.0                        		\\ 
\hline \vspace{0.1cm}
$P(T_0)$ [s]         		        			& 3.7635537(2)				& 3.76363799(7)                    		& 3.763982(3)      		 		\\ \vspace{0.1cm} 
$\dot{P}(T_0)$ [s s$^{-1}$]          		& $6.61(4)\times10^{-12}$		& $1.360(6)\times10^{-11}$                & $2.96(4)\times10^{-11}$   	  	\\ \vspace{0.1cm}
$\ddot{P}$ [s s$^{-2}$]   				& $4(3)\times10^{-19}$		& $3.7(2)\times10^{-19}$                    & $0.67(20)\times10^{-19}$    		 \\ 
\hline \vspace{0.1cm}
$\nu(T_0)$ [Hz]      		        			& 0.265706368(14)			&  0.26570037(5)                    		 & 0.2656761(2)       	  			\\ \vspace{0.1cm}
$\dot{\nu}(T_0)$ [Hz s$^{-1}$]  			& $-4.67(3)\times10^{-13}$	& $-9.60(4)\times10^{-13}$ 	         & $-2.09(3)\times10^{-12}$ 	 	 \\ \vspace{0.1cm}
$\ddot{\nu}$ [Hz s$^{-2}$]        			& $-3(2)\times10^{-20}$		& $-2.6(1)\times10^{-20}$                   & $-0.47(11)\times10^{-20}$   		 \\
\hline  \vspace{0.1cm}
rms residual			        			&  0.15 s					& 0.396 s						 & $3 \times 10^{-5}$ s			\\ \vspace{0.1cm}		
$\chi_\nu^2$ (dof)		        			& 0.85 (5)					& 6.14 (44)   		                		 & 3.16 (18)                       			 \\ 
\hline
\end{tabular}
\label{tab:timing}
\end{table*}

\section{Data analysis}
\label{analysis}

\subsection{Timing analysis}
\label{timing}

An updated timing solution was obtained by adopting the same procedure used to find Solution B reported by Coti Zelati et al. (2015): we measured the spin frequency in each of the observations performed between 2014 February 21 (MJD 56\,709.5) and 2016 July 18 (MJD 56\,588.5) by fitting a linear function to the phases determined over 7 ks-long time intervals, and then fitted a quadratic function to the values of the spin frequency to determine the magnetar spin evolution. In the last 2 observations (in 2016 October), the counting statistics becomes so low to prevent the detection of the spin signal. A search for pulsations by means of a fast Fourier transform of the time series as well as the $Z^2_n$ test (Buccheri et al. 1983) with the number $n$ of the assumed harmonics of the signal being varied from 1 to 3 did not reveal any peak above a confidence level of 3$\sigma$ (estimated accounting for the number of independent frequencies examined and also for the number of harmonics in the case of the $Z^2_n$ test), either in the individual observations or in the merged event lists. The 3$\sigma$ upper limit on the 0.3--8 keV pulsed fraction for any signal in the period range expected from extrapolation of the timing solution is about 45 per cent for both observations, accounting for the background level.

\begin{figure}
\hspace{-0.5cm}
\includegraphics[width=1.15\columnwidth]{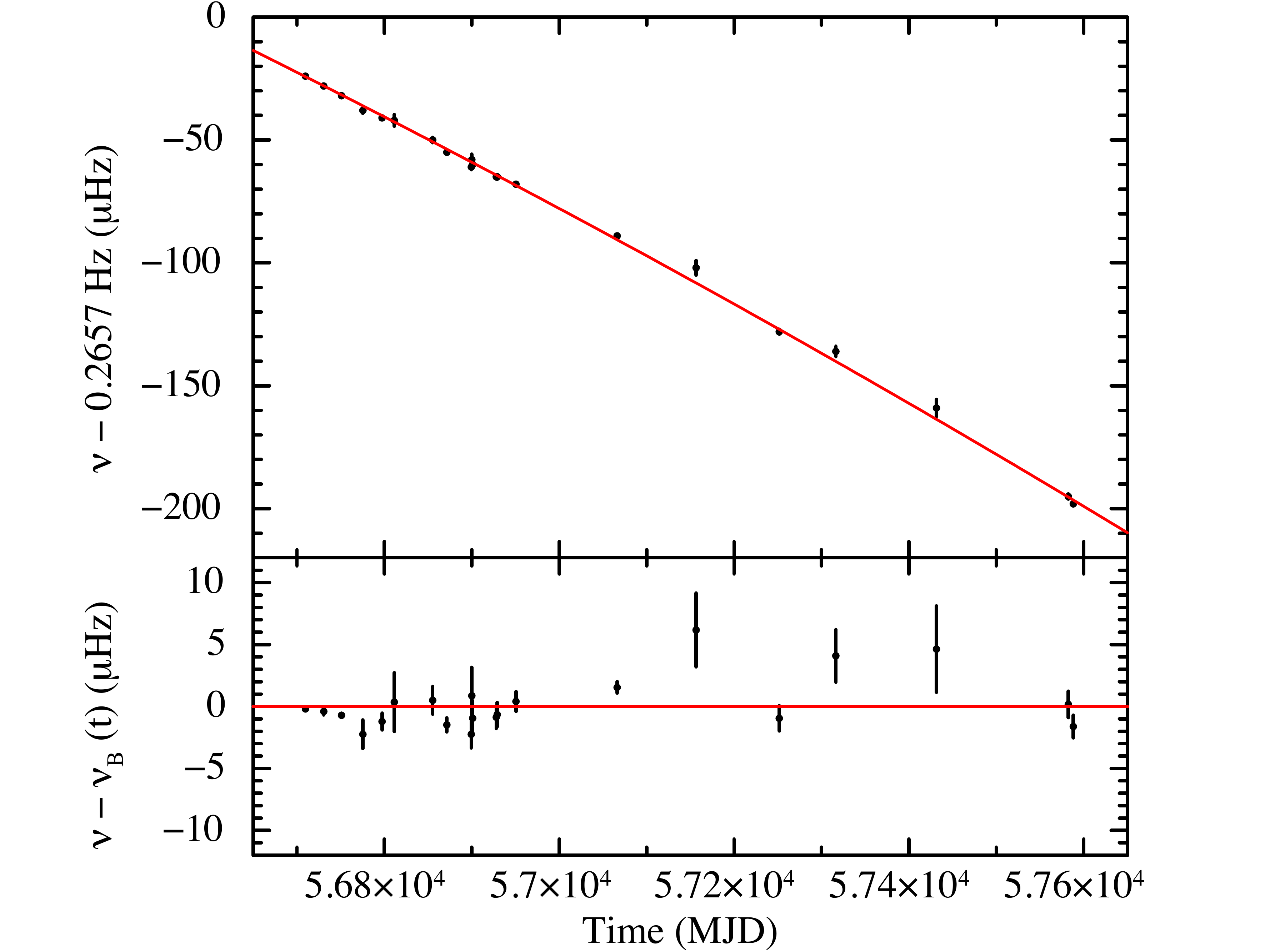}
\vspace{-0.2cm}
\caption{{\em Top panel}: 
temporal evolution of the spin frequency of \galcen\ over the time validity interval MJD 56\,709.5 -- 57\,588.5 (see Figure 1 of Coti Zelati et al. 2015 for the temporal evolution at earlier times). The best-fitting model is marked by the red solid line (see Table~\ref{tab:timing}). {\em Bottom panel}: residuals with respect to the model.}  
\label{fig:timing}
\vskip -0.1truecm
\end{figure}

\begin{figure*}
\begin{center}
\includegraphics[width=17.5cm]{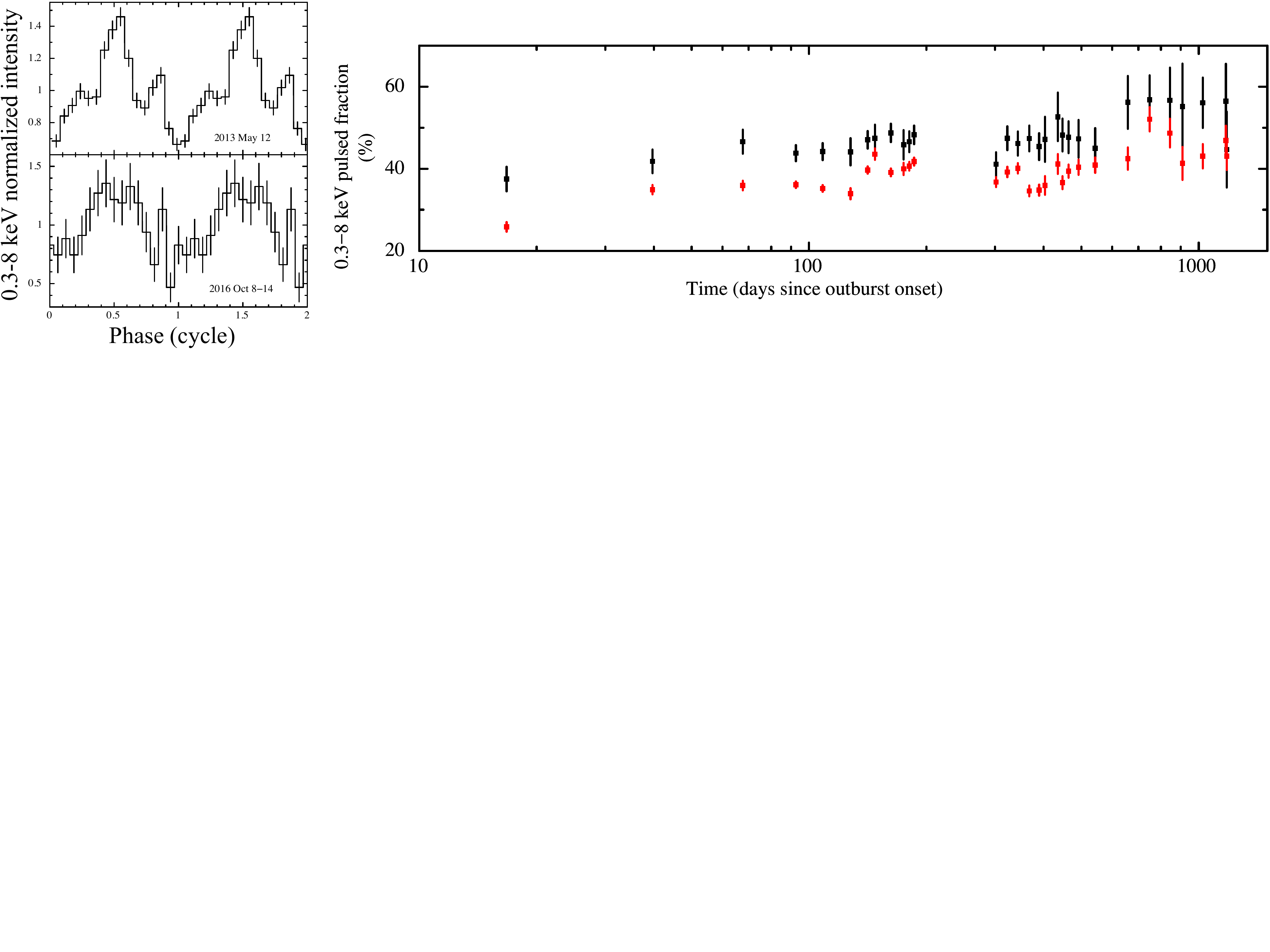}
\vspace{-8.5cm}
\end{center}
\caption{Left-hand panel: 0.3--8~keV background-subtracted and exposure-corrected light curves of \galcen\ folded on the best period for observation IDs 14702 and 18732, and sampled in 16 phase bins. For better visualization, the profiles have been phase-aligned and two cycles are shown. Right-hand panel: temporal evolution of the 0.3--8~keV pulsed fraction. Black points refer to the peak-to-peak pulsed fraction, whereas red points refer to the values obtained from the modeling of the profiles (see text for details). Data points relative to the first 491~d of the outburst were already published by Coti Zelati et al. (2015).} 
\label{fig:efolds}
\vskip -0.1truecm
\end{figure*}

Figure~\ref{fig:timing} shows the overall temporal evolution of the spin frequency from February 2014 to July 2016,  together with the best-fitting model and post-fit residuals. Table~\ref{tab:timing} reports the updated timing solution. The left-hand panel of Figure~\ref{fig:efolds} shows the 0.3--8~keV background-subtracted and exposure-corrected light curves folded on the best period, and sampled in 16 phase bins, for the first (ID 14702; top) and last (ID 18732; bottom) ACIS-S observations where the spin signal is detected. The right-hand panel displays the temporal evolution of the 0.3--8~keV pulsed fraction (PF) of \galcen, derived either as ${\rm PF}=(max-min)\times(max+min)^{-1}$ ($max$ and $min$ refer to the maximum and the minimum of the pulse profile; see black data points in the Figure) or by fitting a constant plus three sinusoidal functions to the pulse profiles, and dividing the value of the semi-amplitude measured for the fundamental sinusoidal component for the average count rate (the sinusoidal periods were fixed to those of the fundamental and harmonic components; see red data points in Figure~\ref{fig:efolds}).
\galcen\ shows a slightly increasing trend in the time evolution of its pulsed fraction, from an initial peak-to-peak value of about $\sim 37$ per cent about 17--18 days after the outburst onset, up to $\sim 55$ per cent after about 600 days.

\begin{table}
\caption{\cxo\ spectral fitting results. Spectra of ACIS-S observations were fitted together with an absorbed black body model ($\chi^2_\nu = 1.01$ for 2506 d.o.f.). The derived absorption column density was $\nh = (1.87\pm0.01) \times 10^{23}$~cm$^{-2}$. Spectra of the grating observations were fitted separately, as described by Coti Zelati et al. (2015). The black body radius and luminosity are evaluated for an observer at infinity and assuming a source distance of 8.3 kpc. Fluxes and luminosities are reported for the 0.3--10~keV energy interval.} 
\begin{threeparttable}
\resizebox{\columnwidth}{!}{
\begin{tabular}{ccccc}
\hline
Obs ID 	     		 & $kT_{BB}$			& $R_{BB}$			& Absorbed flux						& Luminosity	\\
				 & (keV)		   		& (km) 				& (10$^{-12}$~\flux)	& (10$^{35}$~\lum)\\ 
\hline
\vspace{0.08cm}  
14702			  & $0.88\pm0.01$		& 2.52$_{-0.08}^{+0.09}$	& 16.3$_{-0.8}^{+1.0}$				& $4.9\pm 0.5$			\\ \vspace{0.08cm}  
15040\tnote{*}		  & $0.89\pm0.02$		& $2.5\pm0.1$			& 15.5$_{-1.3}^{+0.03}$				& $4.7\pm0.5$			\\ \vspace{0.08cm}
14703	     		  & $0.85\pm0.01$ 		& 2.50$_{-0.08}^{+0.09}$  & 12.7$_{-0.6}^{+0.5}$				& $4.1\pm0.4$			\\ \vspace{0.08cm} 
15651\tnote{*}		  & $0.87\pm0.03$		& $2.4\pm0.2$			& 12.5$_{-0.9}^{+0.07}$				& $3.8\pm0.4$			\\ \vspace{0.08cm}
15654\tnote{*}		  & $0.88\pm0.04$		& $2.4\pm0.2$			& 12.4$_{-0.9}^{+0.05}$				& $3.5\pm0.4$			\\ \vspace{0.08cm}
14946	     		  & $0.83\pm0.01$ 		& 2.39$_{-0.08}^{+0.09}$	& 10.4$_{-0.7}^{+0.4}$				& $3.5\pm0.3$			\\ \vspace{0.08cm} 
15041	     		  & $0.842\pm0.008$ 	& 2.16$_{-0.05}^{+0.06}$	& 9.2$_{-0.3}^{+0.2}$				& 3.0	$_{-0.4}^{+0.2}$	\\ \vspace{0.08cm} 
15042	     		  & $0.834\pm0.008$ 	& $2.09\pm0.05$		& $8.2\pm0.3$						& 2.7$_{-0.4}^{+0.2}$	\\ \vspace{0.08cm} 
14945	     		  & $0.85\pm0.01$ 		& 1.89$_{-0.07}^{+0.08}$  & 7.7$_{-0.4}^{+0.3}$				& $2.4\pm0.2$			\\ \vspace{0.08cm} 
15043	     		  & $0.824\pm0.008$ 	& 2.03$_{-0.05}^{+0.06}$ 	& 7.2$_{-0.3}^{+0.2}$				& 2.4	$_{-0.3}^{+0.2}$	\\ \vspace{0.08cm} 
14944	     		  & $0.84\pm0.01$ 		& 1.88$_{-0.07}^{+0.08}$	& $7.0\pm0.4$						& 2.3$_{-0.3}^{+0.2}$	\\ \vspace{0.08cm} 
15044	     		  & $0.814\pm0.009$		& $1.98\pm0.06$ 		& $6.4\pm0.2$						& 2.2$_{-0.3}^{+0.2}$	\\ \vspace{0.08cm} 
14943	     		  & $0.81\pm0.01$		& 1.95$_{-0.08}^{+0.09}$ 	& 6.1$_{-0.4}^{+0.2}$				& $2.1\pm0.3$			\\ \vspace{0.08cm} 
14704	     		  & $0.810\pm0.009$ 	& $1.94\pm0.06$		& 5.9$_{-0.3}^{+0.2}$				& $2.1\pm0.2$			\\ \vspace{0.08cm} 
15045			  & $0.825\pm0.009$		& $1.83\pm0.05$	 	& 5.9$_{-0.2}^{+0.1}$				& 2.0	$_{-0.2}^{+0.1}$	\\ \vspace{0.08cm}
16508	     		  & $0.82\pm0.01$ 		& $1.49\pm0.05$ 		& $3.7\pm0.1$						& 1.3	$_{-0.2}^{+0.1}$	\\ \vspace{0.08cm}
16211	     		  & $0.81\pm0.01$ 		& $1.50\pm0.05$ 		& 3.4$_{-0.2}^{+0.1}$				& $1.2\pm0.2$			\\ \vspace{0.08cm}
16212			  & $0.82\pm0.01$ 		& $1.38\pm0.05$ 		& 3.1$_{-0.2}^{+0.1}$				& $1.1\pm0.1$			\\ \vspace{0.08cm}
16213			  & $0.81\pm0.01$ 		& $1.37\pm0.05$ 		& $3.0\pm0.1$						& $1.0\pm0.1$			\\ \vspace{0.08cm}
16214			  & $0.81\pm0.01$ 		& $1.34\pm0.05$ 		& $2.7\pm0.4$						& $1.0\pm0.1$			\\ \vspace{0.08cm}
16210			  & $0.84\pm0.02$ 		& 1.17$_{-0.06}^{+0.07}$	& 2.6$_{-0.3}^{+0.1}$				& $0.9\pm0.1$ 			\\ \vspace{0.08cm}
16597			  & $0.77\pm0.02$		& 1.36$_{-0.08}^{+0.09}$	& $2.1\pm0.4$						& $0.8\pm0.1$			\\ \vspace{0.08cm} 
16215			  & $0.81\pm0.01$ 	 	& $1.16\pm0.05$ 		& $2.1\pm0.3$						& $0.73\pm0.08$		\\ \vspace{0.08cm}
16216			  & $0.77\pm0.01$		& 1.27$_{-0.05}^{+0.06}$ & $1.9\pm0.2$						& $0.73\pm0.07$		\\ \vspace{0.08cm}
16217			  & $0.77\pm0.01$		& $1.24\pm0.06$		& $1.8\pm0.2$						& $0.69\pm0.09$		\\ 
\hline
\hline
16218\tnote{$\dagger$}			  & $0.79\pm0.01$		& 1.09$_{-0.05}^{+0.06}$ 	& $1.7\pm0.2$						& $0.60\pm0.07$		\\ \vspace{0.08cm}
16963\tnote{$\dagger$}			  & $0.79\pm0.02$		& 0.98$_{-0.06}^{+0.07}$	& $1.3\pm0.3$						& $0.46\pm0.06$		\\ \vspace{0.08cm}
16966\tnote{$\dagger$}			  & 0.76$_{-0.02}^{+0.03}$ & 0.97$_{-0.08}^{+0.09}$ & $1.0\pm0.2$						& $0.40\pm0.05$		\\ \vspace{0.08cm}
16965\tnote{$\dagger$}			  & $0.72\pm0.02$		& 0.92$_{-0.07}^{+0.09}$	& $0.7\pm0.2$						& $0.29\pm0.04$		\\ \vspace{0.08cm}
16964\tnote{$\dagger$}			  & $0.74\pm0.03$		& 0.79$_{-0.08}^{+0.10}$ 	& $0.6\pm0.2$						& $0.24\pm0.03$		\\ \vspace{0.08cm}
18055\tnote{$\dagger$}			  & $0.71\pm0.04$		& 0.76$_{-0.11}^{+0.15}$	& $0.4\pm0.2$						& $0.18\pm0.03$		\\ \vspace{0.08cm}
18056\tnote{$\dagger$}			  & 0.75$_{-0.04}^{+0.05}$ & 0.68$_{-0.12}^{+0.15}$ & $0.4\pm0.2$						& $0.18\pm0.02$		\\ \vspace{0.08cm}
18731\tnote{$\dagger$}			  & $0.70\pm0.02$		& 0.70$_{-0.06}^{+0.07}$ 	& $0.31\pm0.02$					& $0.15\pm0.02$		\\ \vspace{0.08cm}
18732\tnote{$\dagger$}			  & $0.71\pm0.02$		& 0.72$_{-0.05}^{+0.06}$	& $0.35\pm0.02$					& $0.17\pm0.02$		\\ \vspace{0.08cm}
18057\tnote{$\dagger$}			  & $0.66\pm0.03$ 		& 0.79$_{-0.09}^{+0.12}$  & $0.26\pm0.02$					& $0.14\pm0.02$		\\ \vspace{0.08cm}
18058\tnote{$\dagger$}			  & $0.64\pm0.03$ 		& 0.78$_{-0.10}^{+0.11}$  & $0.23\pm0.02$					& $0.13\pm0.02$		\\ 
\hline
\end{tabular}
}
\begin{tablenotes}
\item[$\dagger$] New unpublished observations.
\end{tablenotes}
\end{threeparttable}
\label{tab:spectralfits}
\end{table}

\begin{figure*}
\begin{center}
\includegraphics[width=2.\columnwidth]{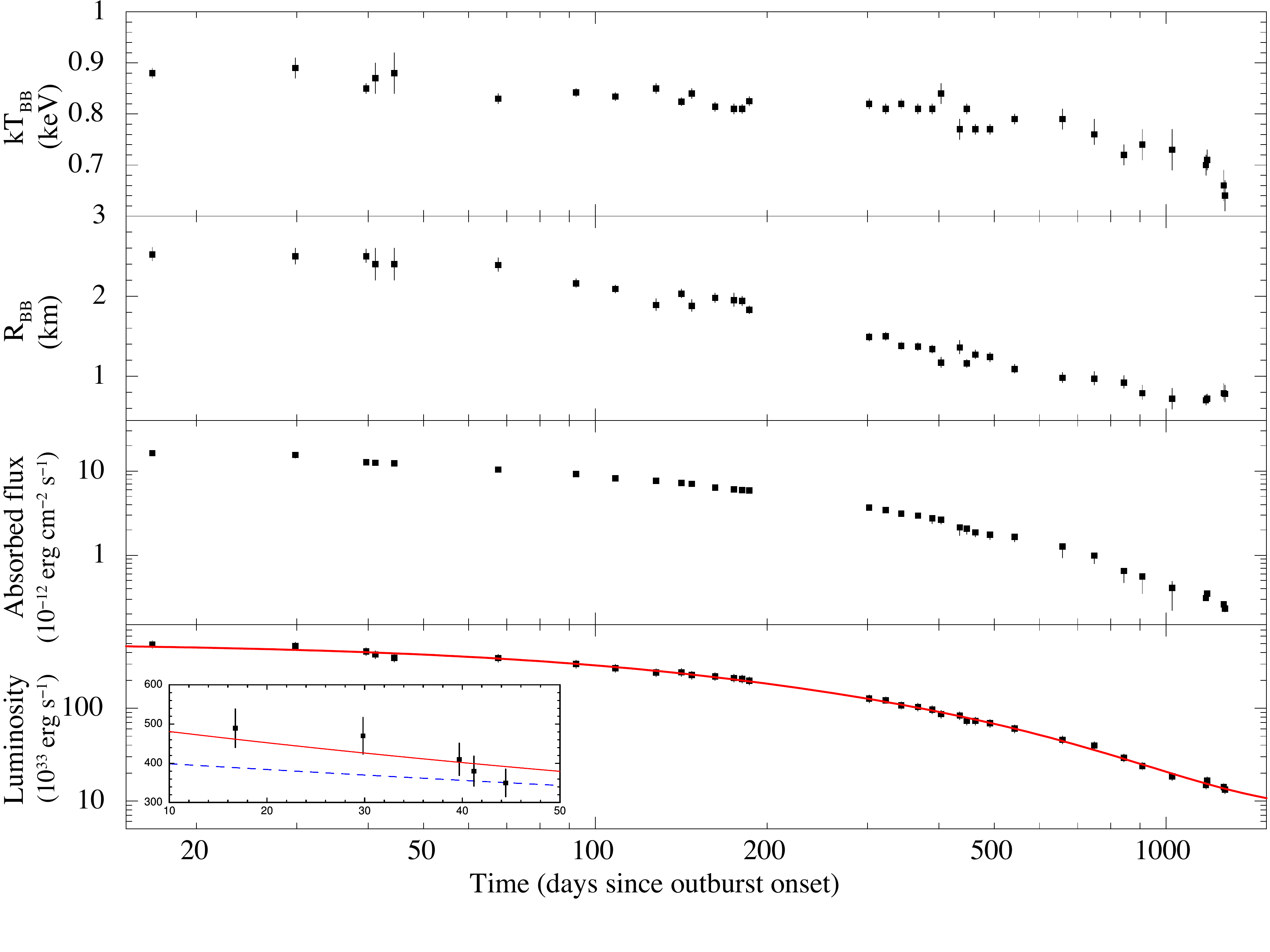}
\vspace{-1cm}
\end{center}
\caption{Time evolution of the black body temperature and radius, and of the 0.3--10~keV absorbed flux and luminosity of \galcen. Spectra of the ACIS-S observations within the first $\sim 200$ days of the outburst were corrected for the effects of pile-up. The best-fitting model for the luminosity decay (two exponential functions plus a constant) is marked by the red solid line in the bottom panel. The double exponential and the single exponential models are compared in the inset of the lower panel for the early stages of the outburst, and are depicted by the red solid line and the blue dashed line, respectively. Data points relative to the first 491~d of the outburst were already published by Coti Zelati et al. (2015).}
\label{fig:parameters}
\end{figure*}

\subsection{Spectral analysis}
\label{spectra}

All background-subtracted spectra were fitted within \textsc{xspec}\footnote{http://heasarc.gsfc.nasa.gov/xanadu/xspec/} (version 12.9.0g; Arnaud 1996). We adopted the Tuebingen-Boulder model (\textsc{tbabs} in \textsc{xspec}), the abundances of Wilms, Allen \& McCray (2000) and the cross-sections of Verner et al. (1996), to account for photo-electric absorption by neutral material in the interstellar medium (ISM) along the line of sight (LoS). The model developed by Davis (2001) was included for the first 12 data sets (obs IDs from 14702 to 15045, corresponding to the first $\sim 200$ days of the outburst decay) to model the spectral distorsions induced by pile-up (see Coti Zelati et al. 2015 for more details). 

\subsubsection{Black body model}
\label{models}

We fitted all spectra together with an absorbed black body model. The hydrogen column density ($\nh$) was tied up across all spectra, as no significant variations were observed over the whole 3.5~yrs of monitoring, whereas the black body temperature and radius were left free to vary (this was indeed the best-fitting model describing the temporal evolution of the 0.3--8~keV spectral shape during the first 1.5~yr; Coti Zelati et al. 2015).

The joint fit yielded $\chi^2_\nu = 1.01$ for 2531 dof. The inferred column density is $\nh = (1.87\pm0.01) \times 10^{23}$~cm$^{-2}$. Table~\ref{tab:spectralfits} reports the best-fitting parameters for the black body component, and Figure~\ref{fig:parameters} shows the temporal evolution of the spectral parameters, the absorbed flux and the luminosity. The source is still fading, and the black body temperature still remains at a relatively high value of $\sim 0.65$~keV about 1270 days after the outburst onset (to be compared with $\sim 0.9$~keV about 2 weeks after the outburst peak). According to this model, the monotonic decline in the X-ray flux is then due mainly to the reduction of the black body radius, which decreased from an initial value of $\sim 2.5$~km down to $\sim 0.8$~km (for an assumed distance of 8.3~kpc and as measured by an observer at infinity; see the left-hand panel of Figure~\ref{fig:lumradius} for the spectra in the first and last ACIS-S observations). We gauged the rate of the hot spot shrinking by fitting an exponential function to the data points, and derived an $e$-folding time $\tau=356_{-22}^{+24}$~d.

\subsubsection{Correction for dust scattering opacity}
\label{dust}

Corrales et al. (2016) have recently noted that the absorption column density values measured towards heavily absorbed sources ($\nh \gtrsim 10^{22}$~cm$^{-2}$) can be largely overestimated if the scattering of X-ray photons on interstellar dust grains is not taken into account properly in the spectral modeling (see also Smith, Valencic \& Corrales 2016). The discrepancy is even more significant for small source extraction regions, because these barely account for the effect induced by dust scattering of removing photons from the LoS and spreading them into a surrounding halo. 

The effects of the dust layer along the LoS towards the low-mass X-ray binary AX\,J1745.6$-$2901 (at only $\sim1.46$ arcmin from \galcen) were recently investigated by Jin et al. (2017), who measured indeed a large alteration of the source spectral shape and flux for small extraction radii. The authors reported that the dust layer is likely located in the Galactic disk a few kpc away from AX\,J1745.6$-$2901, and thus might intervene also along the LoS towards \galcen. As discussed by Ponti et al. (2017), this possibility is corroborated by radio observations of the pulsed emission from \galcen, which led to constrain the position of the obscuring layer along the LoS of the magnetar at a distance of $\sim5.8$ kpc from the Galactic Centre within the spiral arms of the Milky Way (Bower et al. 2014).

\begin{figure}
\includegraphics[width=1.\columnwidth]{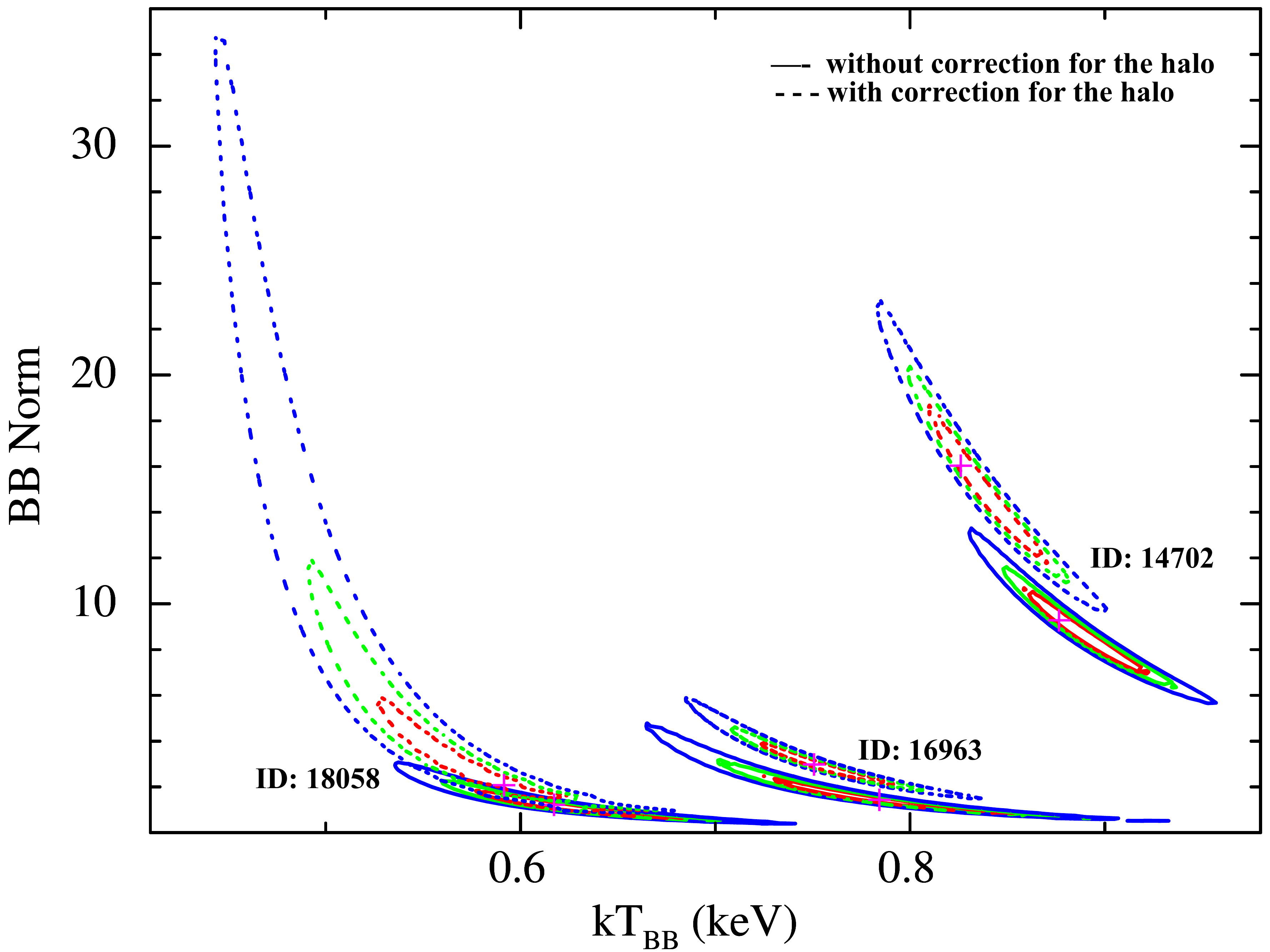}
\vspace{-0.5cm}
\caption{Contour plots in the $kT$--normalization plane for the best-fitting absorbed black body model applied to data sets acquired $\sim 17$, 659 and 1268 d after the outburst onset (obs IDs 14702, 16963 and 18058, respectively). Contour plots derived without and with the correction for the effect of the dust scattering halo are denoted by solid and dashed lines, respectively. Red, green and blue refer to the 1, 2 and 3$\sigma$ confidence intervals, respectively.}
\label{fig:contours1}
\end{figure}

As a further check, we hence decided to correct all magnetar spectra for dust scattering opacity by including in the spectral fits the \textsc{fgcdust} model, developed by Jin et al. (2017) specifically for the case of AX\,J1745.6$-$2901 and applied by Ponti et al. (2017) on a sample of X-ray sources at the Galactic Centre. We derived $\nh = (1.66\pm0.01) \times 10^{23}$~cm$^{-2}$, a value that is a factor of $\sim13$\% lower than that inferred when no correction is applied. Figure~\ref{fig:contours1} shows the variations introduced by this correction to the other spectral parameters. The effective temperature decreases from $\sim0.8$ to $\sim0.6$ keV and the emitting radius shrinks from $\sim3.3$ to $\sim0.9$ km (with $e$-folding time $\tau=295_{-21}^{+24}$~d). The estimated X-ray luminosities are slightly larger compared to those derived without correcting for the dust scattering halo, although we do not detect significant differences in the decay pattern. More sophisticated models are beyond the scope of the present work.

\subsubsection{Depletion of elements in the ISM}

We investigated the X-ray properties of the ISM along the LoS towards \galcen\ by substituting the \textsc{tbabs} model with the \textsc{tbnew} model (Wilms et al. 2000), an improved version where the absorption column densities of different elements are allowed to vary in the fit. We repeated the joint fit leaving the abundances of all elements producing edges in the 1--8 keV energy range free to vary, and with and without the correction for the spectral distorsions caused by the dust scattering halo. The abundances of most elements were found consistent with the Solar values within the uncertainties, and were thus fixed to these values in the following analysis. The only exceptions are represented by Iron and Silicon. Adopting the elemental depletion factors of Wilms et al. (2000) for these elements, for which 30\% and 10\% of the corresponding neutral atoms are expected to be in the gas phase, we computed 3$\sigma$ upper limits well below the Solar abundance values for both elements (see Table~\ref{tab:ironfits}). These results are similar to those reported for the case of the closeby X-ray transient Swift\,J174540.7$-$290015 (Ponti et al. 2016), and are suggestive of depletion of elements into dust grains in the ISM (see Jenkins 2009 for a review about depletion in the ISM).

\subsection{Correlations, outburst energetics, quiescent properties}

\begin{figure*}
\hspace{-0.5cm}
\includegraphics[width=18cm]{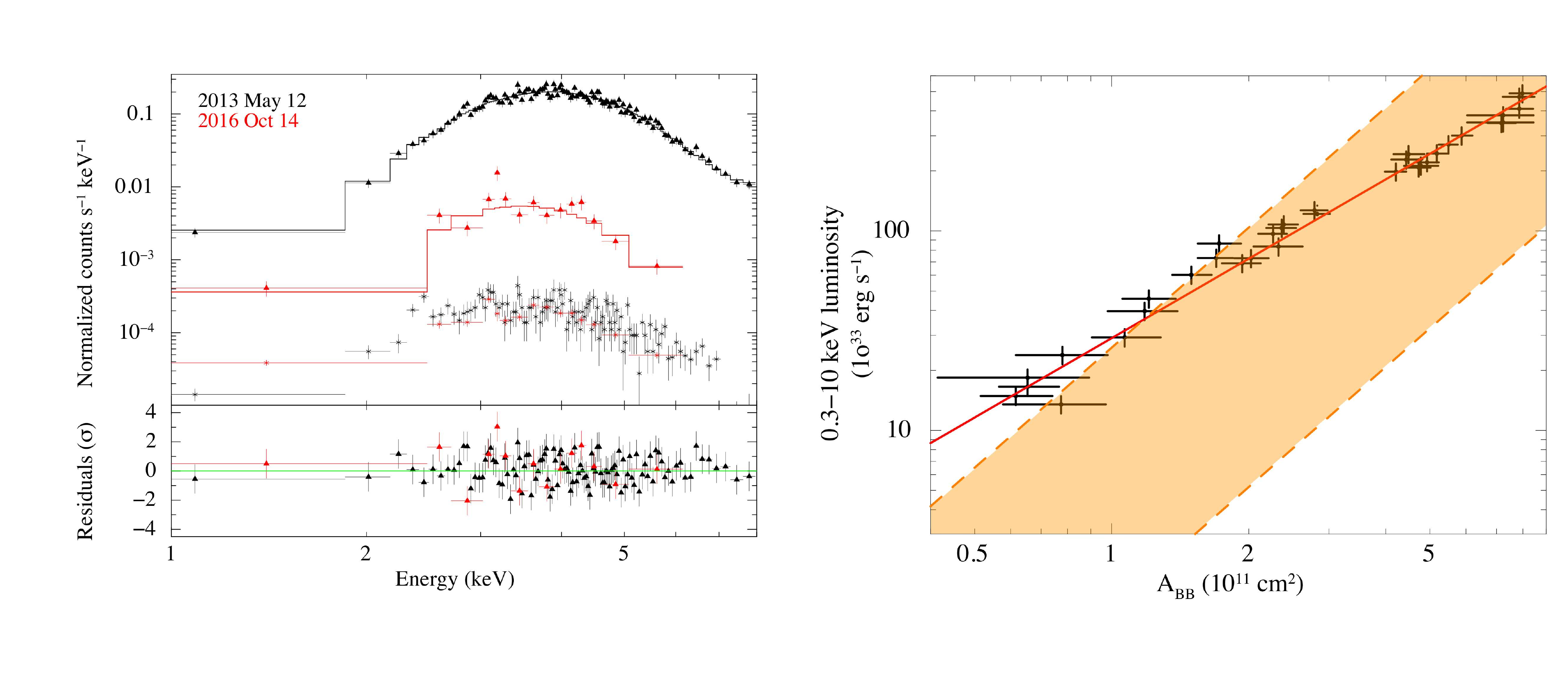}
\vspace{-1.cm}
\caption{Left-hand panel: Background-subtracted spectra of \galcen\ corresponding to the first (black data points; ID: 14702) and last (red data points; ID: 18058) \cxo\ ACIS-S observations.
Background spectra, rescaled to the area of the source extraction region, are also shown. For plotting purpose, spectra are shown in the 1--8~keV after re-binning the data points. The solid lines represent the best-fitting absorbed black body model (see the text for details). Post-fit residuals in units of standard deviations are plotted at the bottom. Right-hand panel: X-ray luminosity of \galcen\ as a function of the black body area. The red solid line denotes the fit with a power law function, the orange dashed lines delimit the region where data points should be located according to the model of Beloborodov (2009; see also Beloborodov \& Li 2016).}
\label{fig:lumradius}
\end{figure*}

The extended monitoring campaign of \galcen\ allows us to search for possible correlations among different parameters across a wide range of values and spectral states. In the context of the black body model for the description of the continuum, we measure a significant correlation between the X-ray luminosity and the area of the thermal hot spot on the neutron star surface ($A_{BB}=4\pi R_{BB}^2$, where $R_{BB}$ is the black body radius inferred from the joint spectral fits; see Section~\ref{models}). The correlation is significant at the 3.1 $\sigma$ level according to the Spearman rank test. To investigate the shape of the correlation, we fitted a power law function of the form $L_X=K\times (A_{BB})^\Gamma$ to the data sets, and found the best-fitting parameters to be $\Gamma = 1.32 \pm 0.02$ and $K = (2.90 \pm 0.08) \times 10^{34}$~erg~s$^{-1}$ cm$^{-2}$ ($\chi^2_\nu = 2.2$ for 31 dof). A completely consistent slope of $\Gamma =1.33  \pm 0.02$ is obtained when all spectra are corrected for the dust scattering opacity along the LoS (see Section~\ref{dust}), although in this case we derived a lower normalization, $K = (1.90 \pm 0.06) \times 10^{34}$~erg~s$^{-1}$ cm$^{-2}$ ($\chi^2_\nu = 3.8$ for 31 dof). The right-hand panel of Figure~\ref{fig:lumradius} shows the 0.3--10~keV X-ray luminosity of \galcen\ as a function of the black body emitting area. The red solid line represents the best-fitting model, whereas the orange lines delimit the region on the $L$ vs. $A$ plane within which data points should lie according to theoretical prediction of the untwisting bundle model for magnetar outbursts (see Figure~12 of Beloborodov \& Li 2016). As discussed by Beloborodov \& Li (2016), a broader range of allowed values on the $L$ vs. $A$ plane may be attained depending on the value of the $B \phi \psi$ term (where $B$ is the magnetic field strength at the surface, $\phi$ is the discharge voltage and $\psi$ is the twist angle of the magnetosphere) and the possible variations of these parameters along the outburst decay.

\begin{table}
\caption{Hydrogen absorption column densities and 3$\sigma$ upper limits on the abundances of Iron (Fe) and Silicon (Si) in the ISM along the LoS towards \galcen. The upper limits were derived from the joint spectral fits of the ACIS-S observations.} 
\resizebox{\columnwidth}{!}{
\begin{tabular}{lccc}
\hline
Model							& $N_{{\rm H}}$		& $A_{{\rm Fe}}$		& $A_{{\rm Si}}$		\\
				 		  		& (10$^{23}$ cm$^{-2}$)	& ($A_{Z,\odot}$) 		& ($A_{Z,\odot}$)			\\ 
\hline
\vspace{0.08cm}  
\textsc{TBnew*bbodyrad}				& $2.30_{-0.04}^{+0.01}$	& < 0.20	& < 0.42		\\
\textsc{fgcdust*TBnew*bbodyrad}		& $2.04_{-0.02}^{+0.01}$	& < 0.19	& < 0.36 	\\
\hline
\end{tabular}
}
\label{tab:ironfits}
\end{table}

\begin{table*}
\caption{Estimates on the quiescent properties of \galcen. The outburst duration is estimated by considering the epoch at which the source attains the quiescent level within an uncertainty of 10\% on the value for the quiescent luminosity, according to the double-exponential model. $\tau_1$ and $\tau_2$ represent the $e$-folding times of the exponential functions, $E$ refer to the estimated energy released, $kT_q$ and $F_{X,q}$ are the upper limits
on the black body temperature and flux detectable in quiescence, respectively. The radius of the black body emitting region was fixed to 10~km. }
\begin{tabular}{cccccccc}
\hline \vspace{0.1cm}	
Assumed $L_{X,q}$ (\lum)	                 &  Outburst duration (days) 	& $\tau_1$ (days) 		& $\tau_2$ (days) 		& $E$ (erg)				& $kT_q$ (keV)		& $F_{X,q}$ (\flux)	& $\chi^2_\nu$ (dof)			\\
\hline \vspace{0.1cm}
$1 \times 10^{32}$				& $\sim 5300$				& 157$_{-26}^{+29}$	  	& 550$_{-65}^{+131}$ 	& $1.1\times 10^{43}$ 		& 0.05			& $5 \times 10^{-15}$  & 0.31 (31) \\ \vspace{0.1cm}
$1 \times 10^{33}$				& $\sim 3800$				& 152$_{-27}^{+30}$  	& 518$_{-57}^{+115}$ 	& $1.1\times 10^{43}$ 		& 0.09 			& $7 \times 10^{-15}$ & 0.31 (31) \\ \vspace{0.1cm}
$1 \times10^{34}$				& $\sim 1700$				& 74$_{-28}^{+33}$ 		& 290$_{-13}^{+16}$ 	& $9.5\times 10^{42}$ 		& 0.17  			& $1 \times 10^{-14}$ & 0.27 (31)  \\
\hline
\end{tabular}
\label{tab:quies_limits}
\end{table*}

The decay of the 0.3--10~keV luminosity can be satisfactorily reproduced by a phenomenological model consisting in a single exponential function plus a constant term: $L_X(t)=A\times \rm{exp}(-t/\tau)$ + $L_q$ ($\chi^2_\nu = 0.81$ for 32 dof). Here $t$ represents the time since the outburst onset (assumed to be coincident with the epoch of detection of the first soft gamma-ray burst from the source on 2013 April 25 at 19:15:25 UT; Kennea et al. 2013), $\tau$ is the e-folding time, and $L_q$ indicates the quiescent level (assumed to be $\lesssim 10^{34}$~\lum, as derived from analysis of all archival \cxo\ observations of the Galactic Centre between 1999 and 2013 April). However, this model does not properly reproduce the very early phases of the outburst decay (within 40 days of the outburst onset; see the blue dashed line in the inset of Figure~\ref{fig:parameters}), which are of paramount importance to adequately estimate the total energy released. A better modelling is obtained when adopting the superposition of two exponential functions: $L_X(t)=\sum_{i=1}^2 A_i\times \rm{exp}(-t/\tau_i)$ + $L_q$ (see the red solid line in Figure~\ref{fig:parameters}). We obtained $\chi^2_\nu = 0.26$ for 30 dof, and determined $\tau_1=96_{-39}^{+48}$~d and $\tau_2=326_{-43}^{+108}$~d. Alternatively, a super-exponential function of the form $L_X(t)=B\times \rm{exp}[-(t/\tau)^\alpha]$ + $L_q$ provides a good description as well ($\chi^2_\nu = 0.24$ for 31 dof), yielding $\tau=162_{-28}^{+26}$~d and $\alpha=0.71\pm0.08$. 

Extrapolation of the double-exponential model between the epoch of the outburst onset and that of the recovery of the quiescent state led to an estimate for the total energy released during the outburst of about $10^{43}$~erg in the 0.3--10~keV energy interval. A fully consistent value is estimated by adopting the super-exponential function to model the luminosity evolution. This value should be considered only as a lower limit, owing to the unknown quiescent level of the source. 

We then computed the outburst energetic with different assumptions on the true quiescent level of the source. Based on the values usually observed (and predicted) for magnetars undergoing major outbursts (Rea \& Esposito 2011; Pons \& Rea 2012; Li \& Beloborodov 2016), we conservatively assume that the source attains a quiescent luminosity in the $10^{32}-10^{34}$~\lum\ range, and consider in particular values of 10$^{32}$, 10$^{33}$ and 10$^{34}$~\lum\ for the following estimates (see Table~\ref{tab:quies_limits}). Table~\ref{tab:quies_limits} reports the outburst total energies and durations obtained always assuming the same double-exponential decay derived above. While the overall duration of the outburst significantly depends on the assumed quiescent flux (a factor of 2.5), the $e$-folding time and the total released energy have a much weaker dependence (only a factor of  $\sim 1.3$ and $\sim 1.1$, respectively), and are in line with the values estimated for other magnetars. 

We also estimated the upper limit on the temperature of \galcen\ detectable during quiescence by assuming that the X-ray thermal emission in this phase will arise from the whole neutron star surface, rather than from hot spots heated along the outburst. We then inferred the limiting value for the quiescent absorbed flux by assuming a black body spectral shape with $\nh = 1.87 \times 10^{23}$~cm$^{-2}$, and the upper limit for the 0.3--8~keV count rate within the error circle of the source during pre-outburst ACIS-S observations performed between 2012 February 6 and October 31 (about 2.9~Ms in total), i.e. $\sim 0.001$ counts s$^{-1}$. All values are reported in Table~\ref{tab:quies_limits}.

\section{Discussion}
\label{discussion}

The extensive X-ray monitoring campaign of the Galactic Centre with \cxo\ allowed us to track the outburst evolution of magnetar \galcen\ over a timespan of $\sim 3.5$ years since its onset, with an unprecedented detail. We measured an increase in the spin-down rate by a factor of $\sim4.5$ along 3 years of outburst decay, from $\dot{P}\sim 6.6\times10^{-12}$~s~s$^{-1}$ up to $\dot{P}\sim3\times10^{-11}$ s s$^{-1}$. Changes in the rotational evolution of magnetars are commonly observed during outbursts. For example, 1E\,1048.1$-$5937 displayed extreme variation in the spin-down rate after its outburst in 2011 December, with the frequency derivative first increasing nearly monotonically by a factor of $\sim4.5-4.6$, and then decaying back to the nominal spin-down value in an oscillatory manner over a timescale of months. Comparison with previous outbursts also suggested that a similar behaviour may repeat on a recurrence time of $\sim1800$ days (Archibald et al. 2015). SGR\,1806$-$20 was also characterized by a significantly erratic behaviour in its timing properties after the giant flare in 2004 December, showing an increase by a factor of $\sim2-3$ in the long term spin-down rate both in 2006 and 2008 (Younes, Kouveliotou \& Kaspi  2015). Extended X-ray coverage of magnetar \xte\ between 2003 and 2014 led Camilo et al. (2016) and Pintore et al. (2016) to investigate the timing properties of the source over a long temporal baseline. The source was characterized by noisy spin-down up to 2007, with non-monotonic excursions in the frequency derivative by a factor $\gtrsim8$, and was then observed to spin-down regularly in the following 7 years.

The evolution of the timing properties of \galcen\ could be compatible with the predictions of the untwisting bundle model for magnetar outbursts (Beloborodov 2009). According to this scenario, the evolution of magnetar outbursts is regulated by the presence of a twisted bundle of current-carrying closed field lines in a confined region of the magnetosphere. As the twist grows during the early stages of the outburst, a larger fraction of closed field lines open out across the light cylinder, yielding an increase of the dipolar magnetic field at the light cylinder radius and hence of the spin-down torque acting on the star (see also Parfrey, Beloborodov \& Hui 2013). 
For a moderate initial twist ($\psi_0<1$~rad), the process leading to the enhancement in the spin-down rate might lag with respect to the epoch of the outburst onset, with a delay comparable to the time required for the twist amplitude to grow from its initial value up to $\psi\sim1$~rad. Conversely, if the initial twist is strong ($\psi_0\gtrsim1$~rad), this mechanism should operate soon after the twist is implanted (Beloborodov 2009). In the former regime, the variation in the spin-down rate can be related to the twist amplitude as 
\be
  2\pi\frac{\Delta\dot{P}}{\dot{P}} \sim \psi^2\ln\frac{u_*}{u_\mathrm{LC}}~,
\en
where $u$ is the area of the bundle calculated at the star surface ($u_*$) and at the light cylinder radius ($u_{{\rm LC}}$). For a strong initial twist, this simple estimate does not hold anymore, and must be substituted by a full non-linear computation. 

The overall fractional increase in the spin-down torque of \galcen, $\Delta\dot{P}/\dot{P} \sim4.5$, appears incompatible with the presence of a moderate twist, according to the formula above. It might be instead the result of prolonged, delayed effects of a strong growing twist during the initial phases of the outburst. However, detailed calculations would be needed to explore this possibility. The spin-down torque is expected to decrease back to that of the dipole field as the twist gradually decays in the subsequent phases of the outburst. A reversal in the rotational evolution should hence be observed at some point in future observations according to this model. The case of \galcen\ may be indeed similar to that of XTE\,J1810$-$197, where a decreasing trend in the spin-down rate was detected only after the first $\gtrsim 500$ days of the outburst (Camilo et al. 2016; Pintore et al. 2016). The ongoing \cxo\ monitoring program of the Galactic Centre will allow us to explore this possibility in \galcen.
Alternatively, the observed torque changes might be ascribed either to the occurrence of (at least two) glitches that the sparse observations could not resolve properly in the timing residuals, or to an increase of the particle density in the magnetosphere possibly related to the outburst activity.

The pulsed fraction showed a slightly increasing trend in time, from an initial peak-to-peak value of $\approx 40$ per cent during the very early stages of the outburst, up to $\approx 55$ per cent about 3 years later (see Figure~\ref{fig:efolds}). A similar behaviour has been observed so far only in the low-field magnetar \lowba, which reached values for the pulsed fraction of $\sim 70-80$ per cent at the quiescent level (see Figure~1 by Rea et al. 2013b). 

All spectra can be successfully described by a black body model, corrected by the large absorption by the neutral material in the ISM along the LoS towards the source ($\nh \sim 1.7 \times 10^{23}$~cm$^{-2}$, comparable with that estimated for other sources located at the Galactic Centre, including \sgras, using the same abundances and cross-sections; see Ponti et al. 2017). The effects of dust scattering in the ISM were also taken into account, yielding only subtle variations in the values of the spectral parameters modelling the continuum emission and of the luminosities. 

\galcen\ is still recovering from the outburst episode that led to its discovery in April 2013. 
The spectral evolution implied by the black body model suggests the presence of a hot spot on the neutron star surface, presumably formed during the episode that led to the outburst activation, and then progressively shrinking and cooling. However, some care must be exercised in extracting physical information from the modeling of the thermal emission with a simple black body spectrum, in particular in assessing the shrinking of the emitting region. The true emission process responsible for the thermal radiation from magnetars is still largely unknown. It may be related to the presence of an atmosphere, albeit with  properties quite different from those of standard atmospheres around passively cooling neutron stars, or even arise from a condensed surface: in both cases the spectrum is expected to be thermal but not necessarily black body-like (see e.g. Potekhin 2014 for a recent review). The fitting with more physically motivated models might return somehow different values of the area and temperature of the emitting region, possibly showing a different trend for the hot spot radius. We have also fitted all the spectral data sets using the \textsc{NTZ} model (Nobili, Turolla \& Zane 2008a,b), which is a three-dimensional implementation of the resonant cyclotron up-scattering of thermal photons coming from the entire star surface (assumed to be at the same temperature) onto electrons flowing in a globally twisted magnetosphere of a magnetar. We obtained an equally statistically acceptable result ($\chi_\nu^2= 0.99$ for 2467 dof) by leaving all parameters free to vary, although no tight constraints could be derived on the values of the bulk motion velocity of the magnetospheric charged particles and the twist angle (as expected given the thermal spectral shape). The estimated absorption column density, $\nh = (1.98\pm0.02) \times 10^{23}$~cm$^{-2}$, is slightly larger compared to that derived from the black body model, whereas the effective temperature, fluxes and luminosities turn out to be fully consistent with the black body values within the uncertainties. These results indicate that the overall magnetar spectral evolution along the outburst decay can be properly described also without necessarily invoking the existence of a shrinking emission region on the neutron star surface.

Taking the black body model as a baseline, the observed properties can be hardly reconciled with crustal cooling models, for which the size of the emitting region is expected to remain fairly constant while the temperature decreases as part of the internal heat is conducted up to the outer layers of the crust (see Coti Zelati et al. 2015 for our previous simulations). On the other hand, the observed behaviour is again a natural prediction of the untwisting bundle model. 
Recent studies have shown that the resistive untwisting of magnetic bundles in magnetar outbursts might last a few years (e.g., Beloborodov \& Li 2016). The prolonged, sustained emission of \galcen\ may be indeed due to the continuous bombardment of returning currents onto the neutron star surface. The gradual dissipation of the twist should lead to the reduction of both the size and temperature of the region at the base of the bundle as the rate of charged particles impacting upon the surface decreases. This is in broad agreement with the shrinking of the hot spot derived by spectral fits (see Figure~\ref{fig:lumradius})\footnote{A power-law like tail associated with resonant cyclotron scattering of thermal photons onto the charged particles in the magnetic twist might be observed in the spectra along the overall decay. For the specific case of \galcen, a faint power-law component was only detected above $\sim 8$ keV, thanks to the broader energy coverage and the larger effective area of the instruments on board \xmm\ (see Coti Zelati et al. 2015) and \nustar\ (see Kaspi et al. 2014).}. 

\begin{figure*}
\hspace{-1cm}
\includegraphics[width=2.0\columnwidth]{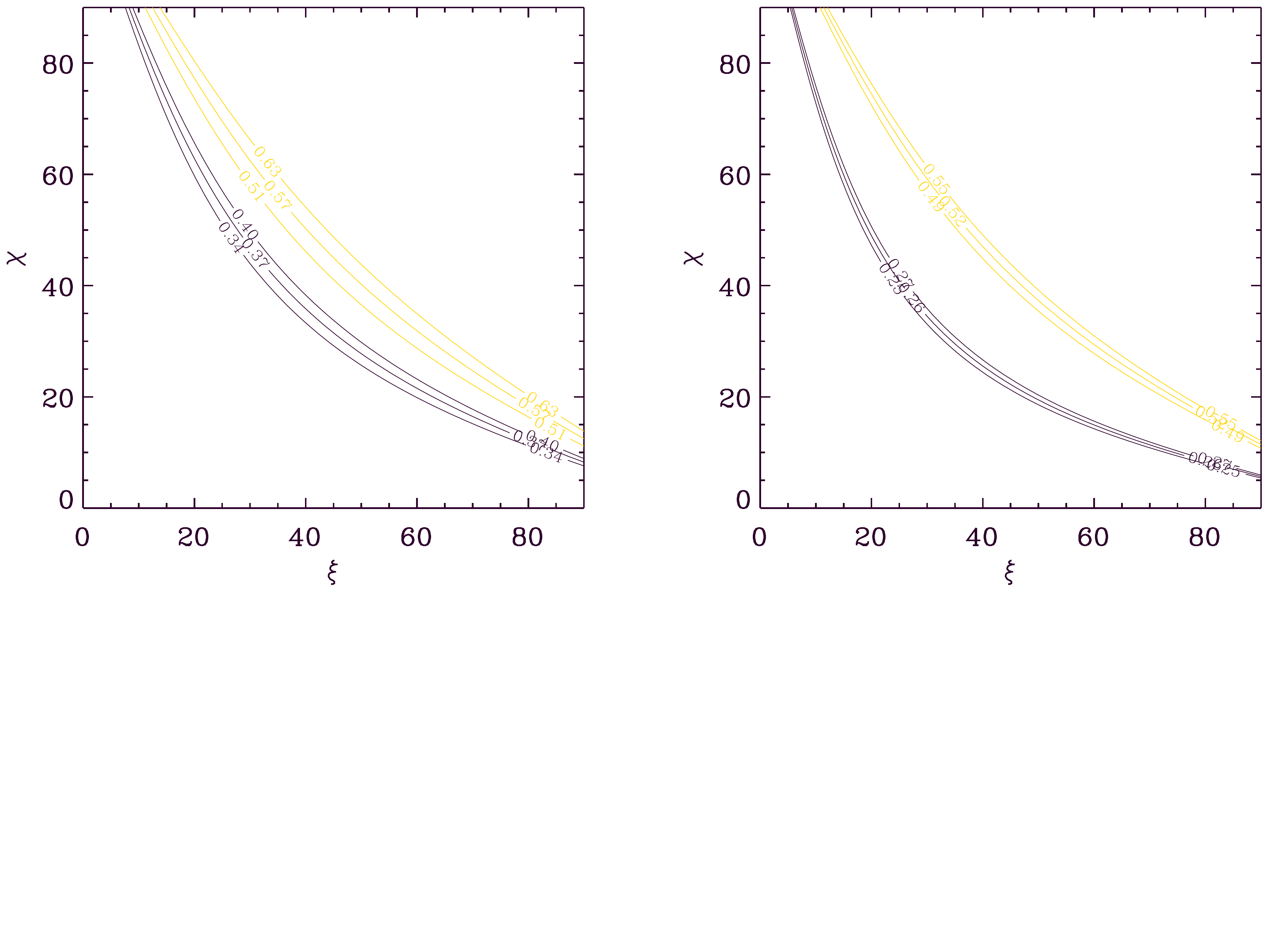}
\vspace{-4.7cm}
\caption{Contour plots for the pulsed fraction, plotted as a function of the viewing angles, for observations performed $\sim 17$ d (ID 14702; black lines) and $\sim 750$ d (ID 16966; yellow lines) after the outburst onset (see the text for details). 1$\sigma$ confidence intervals are reported. The contours in the left-hand panel refer to the peak-to-peak pulsed fractions, whereas those in the right-hand panel refer to the pulsed fractions obtained from the modeling of the pulse profiles (see the text for details).}
\label{fig:angles}
\end{figure*}

Figure~\ref{fig:angles} shows the contour plots of the magnetar pulsed fraction as a function of the viewing angles $\chi$ and $\xi$ (i.e., the inclination of the line-of-sight and the spot normal with respect to the rotation axis, respectively) at different epochs of the outburst decay. For our computations we adopted the analytical expressions derived by Turolla \& Nobili (2013), which account for the effects of gravitational light-bending, and we assumed isotropic black body emission from a finite, uniform, circular cap on the neutron star surface. We set the black body temperature and radius equal to the measured values at the corresponding epochs, and took into account the uncertainties on both parameters (see Table~\ref{tab:spectralfits}). We then calculated the values of $\chi$ and $\xi$ in such a way that the corresponding pulsed fractions were compatible with the observed values within the uncertainties (see the right-hand panel of Figure~\ref{fig:efolds}). No favorable viewing geometry exists that is able to reproduce the observed variations of the pulsed fraction between the two epochs. This is suggestive of an anisotropic emission pattern from the spot, or a change in its position on the surface in time or even a more complex shape of the emitting region with respect to that of a simple circular cap as assumed in our code. However, the latter possibility appears less likely, because modeling the emission with an elongated spot or multiple spots would predict an even smaller pulsed fraction, at odds with the large values observed in the data.

The observations carried out in the past 2 years have confirmed the extremely slow luminosity decay on which we have already reported in our previous study concerning the first year and a half of outburst decay (Coti Zelati et al. 2015). The decay is more rapid in the first 100 days of the outburst ($e$-folding time $\tau_1 \sim 96$~d) than in the following stages ($\tau_2 \sim 330$~d). Prolonged outburst decays have been observed also in other
magnetars. For example, \sgrd\ underwent an outburst in 1998, and several observations with different X-ray instruments revealed that the source returned to its quiescent level in about 10 yr (Esposito et al. 2008). Scholz, Kaspi \& Cumming (2014) reported on X-ray observations of \lowbb\ over the first $\sim 2$ years of the outburst evolution, and showed that the flux decay could be phenomenologically described 
in terms of multiple exponential functions with high decay time-scales (up to about 320 days for a triple-exponential model). More recently, Younes et al. (2015) observed for SGR\,1806$-$20 an exponential decay of the flux 
between 2003 and 2011 with a characteristic timescale of $\sim1.5$~yr. 

For \galcen\ we set a lower limit (owing to the yet unknown quiescent level of the source) on the total energy released in the outburst of $\gtrsim 10^{43}$~erg. Estimates for the outburst energetic, duration and maximum black body temperature detectable during quiescence were also derived under different assumptions for the quiescent level. In particular, according to our estimates, \galcen\ may reach its pre-outburst level either soon at the end of 2017 or on a much longer timescale of about 10 years, depending on the source quiescent luminosity (either $1 \times 10^{34}$ \lum\ or $1 \times 10^{32}$ \lum, respectively). The exquisite spatial resolution of the instruments on board \cxo\ provides us with the unique opportunity to study the properties of this magnetar down to its faintest flux levels, and the ongoing monitoring campaign will allow us to put more stringent constraints on the source quiescent properties and hence on the energy and timescales involved in the outburst.

\section*{Acknowledgements}
The scientific results reported in this article are based on observations obtained with the {\em Chandra X-ray Observatory} and \xmm, an ESA science mission with instruments and contributions directly funded by ESA Member States and and the National Aeronautics and Space Administration (NASA). This research has made extensive use of software provided by the \cxo\ X-ray Center (CXC, operated for and on behalf of NASA by the Smithsonian Astrophysical Observatory (SAO) under contract NAS8--03060) in the application package \textsc{ciao}, and of softwares and tools provided by the High Energy Astrophysics Science Archive Research Center (HEASARC), which is a service of the Astrophysics Science Division at NASA/GSFC and the High Energy Astrophysics Division of the Smithsonian Astrophysical Observatory. FCZ, NR, PE, AB and JE acknowledge funding in the framework of a Vidi award A.2320.0076 of the Netherlands Organization for Scientific Research (PI: N. Rea), and the European COST Action MP1304 (NewCOMPSTAR). FCZ, NR and DFT are supported by grant AYA2015-71042-P. NR and DFT are also supported by grant SGR2014-1073. RT, AT and SM acknowledge financial contribution from the agreement ASI/INAF I/037/12/0 and from PRIN INAF 2014. JAP acknowledges support by grants AYA2015-66899-C2-2-P and PROMETEOII-2014-069. AP is supported via an EU Marie Sk\"odowska-Curie Individual fellowship under contract no. 660657-TMSP-H2020-MSCA-IF-2014. RPM acknowledges financial support from an `Occhialini Fellowship'.
DH acknowledges support from a Natural Sciences and Engineering Research Council of Canada Discovery Grant and a Fonds de recherche du Qu\'{e}bec Nature et Technologies Nouveaux Chercheurs Grant. GP acknowledges support via the Bundesministerium f\"{u}r Wirtschaft und Technologie/Deutsches Zentrum f\"{u}r Luft-und Raumfahrt (BMWI/DLR, FKZ 50 OR 1604) and the Max Planck Society. FCZ acknowledges Chichuan Jin for providing the \cxo\ ACIS-S version of the dust scattering model. We thank Giovanni Fazio, Joseph Hora, Gordon Garmire and Steven Willner for sharing their data and the referee for helpful comments.

\label{lastpage}

\end{document}